# Uncertainty-Aware Semi-supervised Method using Large Unlabelled and Limited Labeled COVID-19 Data

Roohallah Alizadehsani[1,*] (Member, IEEE), Danial Sharifrazi[2], Navid Hoseini Izadi[3], Javad Hassannataj Joloudari[4], Afshin Shoeibi[5,6], Juan M. Gorriz[7], Sadiq Hussain[8], Juan E. Arco[7], Zahra Alizadeh Sani[9,10], Fahime Khozeimeh[1], Abbas Khosravi[1] (Senior Member, IEEE), Saeid Nahavandi[1] (Fellow, IEEE), Sheikh Mohammed Shariful Islam[11,12,13], U Rajendra Acharya[14,15,16] (Senior Member, IEEE)

[1]Institute for Intelligent Systems Research and Innovations (IISRI), Deakin University, Geelong, Australia
[2] Department of Computer Engineering, School of Technical and Engineering, Shiraz Branch, Islamic Azad University, Shiraz, Iran
[3] Dept. of Electrical and Computer Engineering, Isfahan University of Technology, Isfahan, Iran.
[4] Department of Computer Engineering, Faculty of Engineering, University of Birjand, Birjand, Iran
[5] Computer Engineering Department, Ferdowsi University of Mashhad, Mashhad, Iran.
[6] Faculty of Electrical and Computer Engineering, Biomedical Data Acquisition Lab, K. N. Toosi University of Technology, Tehran, Iran.
[7] Department of Signal Theory, Networking and Communications, Universidad de Granada
[8] System Administrator, Dibrugarh University, Assam 786004, India
[9]Rajaie Cardiovascular Medical and Research Center, Iran University of Medical Sciences, Tehran, Iran
[10]Omid hospital, Iran University of Medical Sciences, Tehran, Iran
[11] Institute for Physical Activity and Nutrition, Deakin University, Melbourne, Australia
[12]Cardiovascular Division, The George Institute for Global Health, Australia
[13] Sydney Medical School, University of Sydney, Australia
[14] Department of Electronics and Computer Engineering, Ngee Ann Polytechnic, Singapore,
[15]Department of Biomedical Engineering, School of Science and Technology, Singapore University of Social Sciences, Singapore,
[16]Department of Bioinformatics and Medical Engineering, Asia University, Taiwan.
* Corresponding author: Roohallah Alizadehsani, Institute for Intelligent Systems Research and Innovations (IISRI), Deakin University, Geelong, Australia.
E-mail: ralizadehsani@deakin.edu.au

## Abstract

The new coronavirus has caused more than 1 million deaths and continues to spread rapidly. This virus targets the lungs, causing respiratory distress which can be mild or severe. The X-ray or computed tomography (CT) images of lungs can reveal whether the patient is infected with COVID-19 or not. Many researchers are trying to improve COVID-19 detection using artificial intelligence. In this paper, relying on Generative Adversarial Networks (GAN), we propose a Semi-supervised Classification using Limited Labelled Data (SCLLD) for automated COVID-19 detection. Our motivation is to develop learning method which can cope with scenarios that preparing labelled data is time consuming or expensive. We further improved the detection accuracy of the proposed method by applying Sobel edge detection. The GAN discriminator output is a probability value which is used for classification in this work. The proposed system is trained using 10,000 CT scans collected from Omid hospital. Also, we validate our system using the public dataset. The proposed method is compared with other state of the art supervised methods such as Gaussian processes. To the best of our knowledge, this is the first time a COVID-19 semi-supervised detection method is presented. Our method is capable of learning from a mixture of limited labelled and unlabelled data where supervised learners fail due to lack of sufficient amount of labelled data. Our semi-supervised training method significantly outperforms the supervised training of Convolutional Neural Network (CNN) in case labelled training data is scarce. Our method has achieved an accuracy of 99.60%, sensitivity of 99.39%, and specificity of 99.80% where CNN (trained supervised) has achieved an accuracy of 69.87%, sensitivity of 94%, and specificity of 46.40%.

**Keywords:** Semi-supervised learning, Generative Adversarial Networks, COVID-19, supervised learning, deep learning.

# 1 Introduction

Since the beginning of 2020, the coronavirus disease 2019 (COVID-19) has been spreading all over the globe as it is contagious in an unprecedented manner [1]. In severe cases, it may lead to multiple organ failure, acute respiratory distress, heart problems, secondary infections in a comparatively high fraction of patients, and thus may cause deaths [2, 3]. On January 30th, 2020, The World Health Organization (WHO), announced the outbreak as a "public health emergency of international concern" (PHEIC). The epicentre of the outbreak was the Huanan Seafood Wholesale Market in Wuhan City, Hubei Province, China, and on March 11$^{th}$, 2020, the WHO declared the COVID-19 a pandemic. Early detection and initiation of treatment in severe cases are vital to deal with the disease and thus mitigating mortality [4].

Reverse-transcription polymerase chain reaction (RT-PCR) is generally utilized to confirm COVID-19. The sensitivity and specificity of RT-PCR were not recorded as robust enough for the treatment of the presumptive patients and early detection [5, 6]. A non-invasive imaging technique called computed tomography (CT) scans can play a crucial role in identifying some characteristic symptoms in the lung related to COVID-19 [7, 8]. A study using chest CT scans achieved 97% sensitivity (580/601 patients, 95% CI, 95-98%) for COVID-19 detection [6]. Hence, chest CT scans may be effectively utilized in diagnosis and early detection of the disease. Since COVID-19 causes pulmonary changes, chest CT scans may exhibit imaging features analogous to other types of pneumonia, leading to confusion during COVID-19 detection. Hence, despite the advantages of CT, it is challenging to discriminate COVID-19 from other types of pneumonia.

Accessibility to huge datasets facilitates deep learning tools and applications to process large amount of unstructured information, allowing high level abstractions with enhanced generalisability and permitting key feature extractions [9]. In the domain of medical imaging, deep learning has shown outstanding performance in automatic feature extraction [10-12]. Due to the noteworthy performance that deep learning methods achieved especially in image processing tasks, it is remarkably useful as feature learners or automatic regressors in addition to classification tasks [9]. Deep learning is utilized to differentiate and detect viral and bacterial pneumonia in pediatric chest CT scans [13]. Different imaging features of chest CT can also be detected using deep learning [14, 15]. Chest CT scans of the positive COVID-19 patients have a discrete radiographic pattern: multifocal patchy consolidation, ground-glass opacities, and/or interstitial changes with a predominantly peripheral distribution [6, 7]. Chest CT scans with the help of deep learning methods have illustrated their efficacy in distinguishing COVID-19 from other types of viral pneumonia and thus becomes a useful diagnostic tool. This, in turn, leads to controlling and managing the pandemic situation [16].

Chest X-ray and CT scans have the potential to detect COVID-19 and isolate the patients in time. As most hospitals are equipped with X-ray, it is the first choice of the radiologists. However, chest X-ray images cannot distinguish soft tissues accurately [17]. Chest CT scan can be utilized as an alternative method. As the number of radiologists is scarce and also busy in pandemic situations, automatic detection of COVID-19 from chest images is highly desirable. Utilizing deep learning, Li et al. [18] designed a COVID-19 detection model called COVNet by extracting features from chest CT. Other non-pneumonia and community-acquired pneumonia (CAP) CT exams are conducted to evaluate the robustness of the model. Their model could discriminate CAP from other lung diseases, accurately. Wang et al. [19] utilized the deep learning strategies to derive the graphical features from the CT images of COVID-19 patients. In these images, there are radiographical changes in the case of infected patients. Gozes et al. [20] proposed an automated approach using CT images for quantification, detection, and monitoring of COVID-19 patients. They used robust 3D and 2D models based on deep learning concepts and integrated them with clinical perceptive. They utilized a multi-centre international dataset and generated a corona score using a 3D volume review from thoracic CT features. This score helped the system to compute the evolution of ailment over time.

Several studies have applied new approaches to detect COVID-19 cases using machine learning and deep learning approaches. Hemdan et al. [21] devised a deep learning-based framework dubbed as COVIDX-Net to aid clinicians to detect COVID-19 from X-ray images. They partitioned the dataset into 80% for training and 20% for testing. Zhang et al. [22] presented an anomaly detection deep technique for reliable and efficient COVID-19 detection. Apostolopoulos et al. [23] analysed the X-ray images of normal incidents, confirmed COVID-19 cases, and common bacterial pneumonia for automated detection of COVID-19 patients. They applied a convolutional neural network (CNN) with transfer learning. Their model derived biomarkers related to the COVID-19 illness. Butt et al. [24] compared multiple CNN models and devised a deep learning model based on 3D and 2D networks to classify no-infection, influenza viral pneumonia, and COVID-19 samples. Their model successfully differentiated the non-coronavirus and coronavirus cases per thoracic CT records.

Other approaches include Song et al. [25] who devised a CT diagnosis system based on deep learning technology and termed it as DeepPneumonia for the identification of COVID-19 patients. Main lesion features, including ground-glass opacity (GGO) which are the key in the diagnosis, are located by their model. Sethy et al. [26] proposed a deep learning technique based on X-ray radiographs analysis for detection of COVID-19 patients. They implemented support vector machine classifier using the deep features to discriminate COVID-19 X-ray images from others. Their technique could assist the radiologists in the diagnosis of COVID-19 patients.

The test kits for COVID-19 are limited in hospitals owing to the exponential growth of the cases. Therefore, it is crucial to search for a fast alternative to detect such cases in order to confine the spread. Narin et al. [27] presented detection of COVID-19 patients based on Inception-ResNetV2, InceptionV3, and ResNet50 utilizing chest X-ray images. Confusion matrices and ROC analyses are performed with 5-fold cross-validation. Shi et al. [28] established a deep learning oriented CT and clinical features based prognosis model for assessing the severity of COVID-19 infection. They applied the least absolute shrinkage and selection operator (LASSO), and developed the pneumonia severity index (PSI). Severe patients had higher PSI ($p<0.001$), percentage of infection ($POI_{CT}$), and mass of infection ($MOI_{CT}$) than non-severe ones. Their model proved its efficacy in the prediction of patients' severity.

Pervasive demands have arisen to combat COVID-19 pandemic crisis by designing an automated and efficient diagnosis system. Maghdid et al. [29] presented an accurate deep learning tool with fast detection mechanism for COVID-19 cases. Various CT and X-ray images have been integrated to provide a comprehensive and publicly available dataset. The detection technique consists of transfer learning and deep learning. The network is trained using AlexNet and CNN models on the CT and X-rays dataset. The overview of published researches on COVID-19 detection is presented in Table 1.

Table 1. Summary of works conducted on the detection of COVID-19 patients using deep learning (DL) techniques.

| Researches | Modalities | Number of Cases/images/datasets | DL architecture |
|---|---|---|---|
| Li et al. [18] | CT | 4356 chest CT | Resnet50 |
| Wang et al. [19] | CT | 453 CT images | CNN |
| Gozes et al. [20] | CT | 157 cases | Resnet-50 |
| Hemdan et al. [21] | X-ray | 50 Chest X-ray images | Google MobileNet and modified VGG19 |
| Zhang et al. [22] | X-ray | X-VIRAL and XCOVID datasets | Residual CNN |

| Apostolopoulos et al. [23] | X-ray | 1427 X-ray images | CNN with Transfer Learning |
|---|---|---|---|
| Butt et al. [24] | CT | 219 cases | CNN |
| Song et al. [25] | CT | 275 cases | Resnet50 |
| Sethy et al. [26] | X-ray | 2 datasets | SVM plus Resnet50 |
| Narin et al. [27] | X-ray | 100 images | Inception-ResNetV2, InceptionV3 and ResNet50 |
| Shi et al. [28] | CT | 196 cases | VB-Net |
| Maghdid et al. [29] | CT and X-ray | 170 X-ray images and 361 CT images | AlexNet |
| Zhao et al. [30] | CT | 349 cases | DenseNet-169<br>ResNet-50 |
| Ko et al. [31] | CT | 3993 chest CT images | ResNet-50 |
| Jaiswal et al. [16] | CT | 2492 CT scans | DenseNet201 |
| Ardakani et al. [32] | CT | 1020 CT | Xception<br>ResNet-101 |
| Kumar et al. [33] | CT | 34,006 CT scan slices | Capsule Network |
| Ni et al. [34] | CT | 14,435 participants with chest CT images | 3D U-Net<br>MVP-Net |
| Alom et al. [35] | CT and X-ray | 420 CT<br>704 chest X-ray | RNN with transfer learning |
| Javaheri et al. [36] | CT | 89,145 CT Slices | 3D CNN |
| Saeedi et al. [37] | CT | 349 cases | DenseNet121, ResNet50 V1&V2, InceptionV3 and MobileNet V1&V2 |
| Zhang et al. [38] | CT | 640 images | CNN |
| Wang et al. [39] | CT | 640 images | Graph convolutional network (GCN) and CNN |

Deep neural networks have great representation power but, their performance heavily relies on availability of labelled training data. In case the labelled data is limited, the deep networks won't be able to learn well. However, we can still train the deep networks well by exploiting unlabelled data. The justification behind our hypothesis is the success of transfer learning [40]. In a nutshell, in transfer learning a learner trained for a specific task $t_i$ is modified for another task $t_j$ which bears some similarity to $t_i$. Although tasks $t_i$ and $t_j$ are not the same, the learner trained for $t_i$ can still benefit the similarity between $t_i$ and $t_j$ which accelerates the learning of $t_j$. This is the motivation for our two-phase semi-supervised approach. In the first phase, the *discriminator* is trained to detect valid CT scan images. In the second phase, using the gained expertise from the first phase, the trained *discriminator* can learn sick and healthy CT images faster. The learning boost is due to the fact that regardless of being COVID or healthy, each training/test image is a valid CT scan which the *discriminator* has mastered using unlabelled data.

Considering the above argument, one may be tempted to use non-deep learners in an attempt to reduce the required amount of the training data. However, the major drawback of such learners is that they treat the input samples as vectors. To feed images to such non-deep learners, we are forced to reshape the images into vectors. The reshaping operation destroys the meaningful features that each pixel has to offer relative to its neighbouring pixels. Hence, the application of deep learning based models seems to be inevitable if the features present in image inputs are to be captured properly. Hence using semi-supervised learning to train deep models with a mixture of unlabelled and limited labelled data is the best solution.

The main contribution of our work is twofold. First, we collected a dataset of lung CT scan images useful for training/evaluation of COVID-19 detection methods. Second, to the best of our knowledge, we are the first group to propose the semi-supervised COVID-19 detection method based on generative adversarial network (GAN) [41] to detect this disease. The proposed method has been improved using Sobel edge detection. Despite being semi-supervised, our method is competitive to its supervised counterparts. This feature is beneficial when labelled data is hard to get. Although we have focused on COVID-19 detection, our method is not limited to any specific dataset. The rest of the paper is organized as follows: Section 2 provides prerequisites, Section 3 describes our dataset, Section 4 elaborates the proposed method, Section 5 presents the experimental results and Section 6 concludes the paper.

# 2 Prerequisites

In this section, the required mathematical concept are briefly reviewed. First, GAN is reviewed since our method is based on it. The Gaussian Process is also reviewed since it is used during our experiments.

## 2.1 Generative Adversarial Networks

Originally proposed by GoodFellow, GAN is a generative model with massive applications. Compared to its predecessors, GAN is capable of generating high-quality images which are vivid and sharp. Basically, GAN is made of two neural networks, namely Generator (G) and Discriminator (D). The job of the Generator is to produce high-quality images which are called fake samples. The Discriminator must distinguish between the real and fake samples. The two networks compete against each other in a minimax game. That is why we call them adversarial networks. The objective function based on which the two networks are trained is given as [41]:

$$\min_G \max_D V(D,G) = \mathbb{E}_{x\sim p_{data}(x)}[logD(x)] + \mathbb{E}_{z\sim p_z(z)}\left[\log\left(1 - D\big(G(z)\big)\right)\right], \tag{1}$$

where $p_{data}(x)$ is the real data (available dataset) distribution, $p_z(z)$ is the Generator input noise distribution, D(x) is the Discriminator output, z is the (Gaussian) noise vector and G(z) is the Generator output. As can be seen, the Generator is trying to minimize the objective function in equation (1), while Discriminator is trying to maximize it. To this end, the following gradients are calculated:

$$\nabla_{\theta_d} \frac{1}{m}\sum_{i=1}^{m}\left[logD\big(x^{(i)}\big) + \log\left(1 - D\left(G\big(z^{(i)}\big)\right)\right)\right], \tag{2}$$

$$\nabla_{\theta_g} \frac{1}{m}\sum_{i=1}^{m}\left[\log\left(1 - D\left(G\big(z^{(i)}\big)\right)\right)\right], \tag{3}$$

where $\theta_d$ and $\theta_g$ are the Discriminator and Generator parameters, respectively [41]. The gradients are computed over mini-batch of m samples.

## 2.2 Gaussian Process

Gaussian Process (GP) is a non-parametric probabilistic model which can be used for regression or classification. GP can be considered as an infinite-dimensional Gaussian distribution which is defined on functions as follows [42]:

**Definition:** f is a Gaussian process if for any index set $\{t^{(i)} \in \mathbb{R}^D, i = 1, \dots, n\}$, vector $f(t) = \left[f\big(t^{(1)}\big), \dots, f\big(t^{(n)}\big)\right]^T$ has a multivariate Gaussian distribution of the form $f(t) \sim \mathcal{N}(m(t), K(t,t))$. Each $f\big(t^{(i)}\big)$ is a random variable, $m(t)$ is the mean function and $K(t,t)$ is the covariance matrix of the Gaussian distribution [42]. The mean function is initialized to constant zero ($m(t) = 0$). Each element of $K(t,t)$ is the output of a positive definite kernel function $k: \mathbb{R}^D \times \mathbb{R}^D \to \mathbb{R}$ which receives $t^{(i)}$ and $t^{(j)}$ as input. In this paper we use squared exponential kernel:

$$k\left(t^{(i)}, t^{(j)}\right) = \sigma_f^2 \exp\left\{\sum_{d=1}^{D} \frac{-1}{2l_d^2}\left(t_d^{(i)} - t_d^{(j)}\right)^2\right\} + \sigma_n^2 \delta_{ij}, \quad \delta_{ij} = \begin{cases} 1, & if\ i = j \\ 0, & o.w \end{cases}, \tag{4}$$

where the kernel function hyper-parameters $\sigma_f$, $\sigma_n$, and $l$ are signal variance, noise variance, and length scale, respectively. The hyper-parameters are learned based on the available training data.

### 2.2.1 GP classification

In this paper, we focus on binary classification of healthy and sick people. Hence, in this section, classification using GP is briefly reviewed. Following the binary classification conventions, the two class labels are denoted as$\{-1, +1\}$. The basic idea behind binary classification using GP is as follows. A prior GP is placed over the latent function $f(x)$. Output of GP is squashed through a logistic function $\sigma(.)$ to obtain prior $\pi(x) \triangleq p(y = +1|x) = \sigma(f(x))$. Here $f(x)$ values are not observable and we are not interested in them either. We only care about input vector $x$ and desired class label $y$. The sole purpose of $f(x)$ is to make the model formulation more convenient.

Assuming that the training inputs are aggregated as column vectors in matrix $X$, and their corresponding labels are expressed as vector $y$, the GP inference is carried out in two steps:

1. For a test case $x_*$, the distribution of the latent variable ($f_*$) is computed as [42]:

$$p(f_*|X, y, x_*) = \int p(f_*|X, x_*, f)p(f|X, y)df, \tag{5}$$

   where,

$$p(f|X, y) = \frac{p(y|f)p(f|X)}{p(y|X)}. \tag{6}$$

2. Now $p(f_*|X, y, x_*)$ is used to produce a probabilistic prediction as:

$$\tilde{\pi}_* \triangleq p(y_* = +1|X, y, x_*) = \int \sigma(f_*)p(f_*|X, y, x_*)df_*. \tag{7}$$

## 3 Dataset description

In this study, 10000 lung CT scan images were captured at Omid Hospital in Tehran. The images have been collected from February 2020 to April 2020. The mean and standard deviation of the patients' age were 49.5±18.5 years old with 45% of the cases were male. Each image has been checked by three radiologists to determine whether the case is COVID-19 or not. Ethical approval to conduct use these data was obtained from the hospital. Typical normal and COVID-19 CT images used for this study is shown in Figure 1.

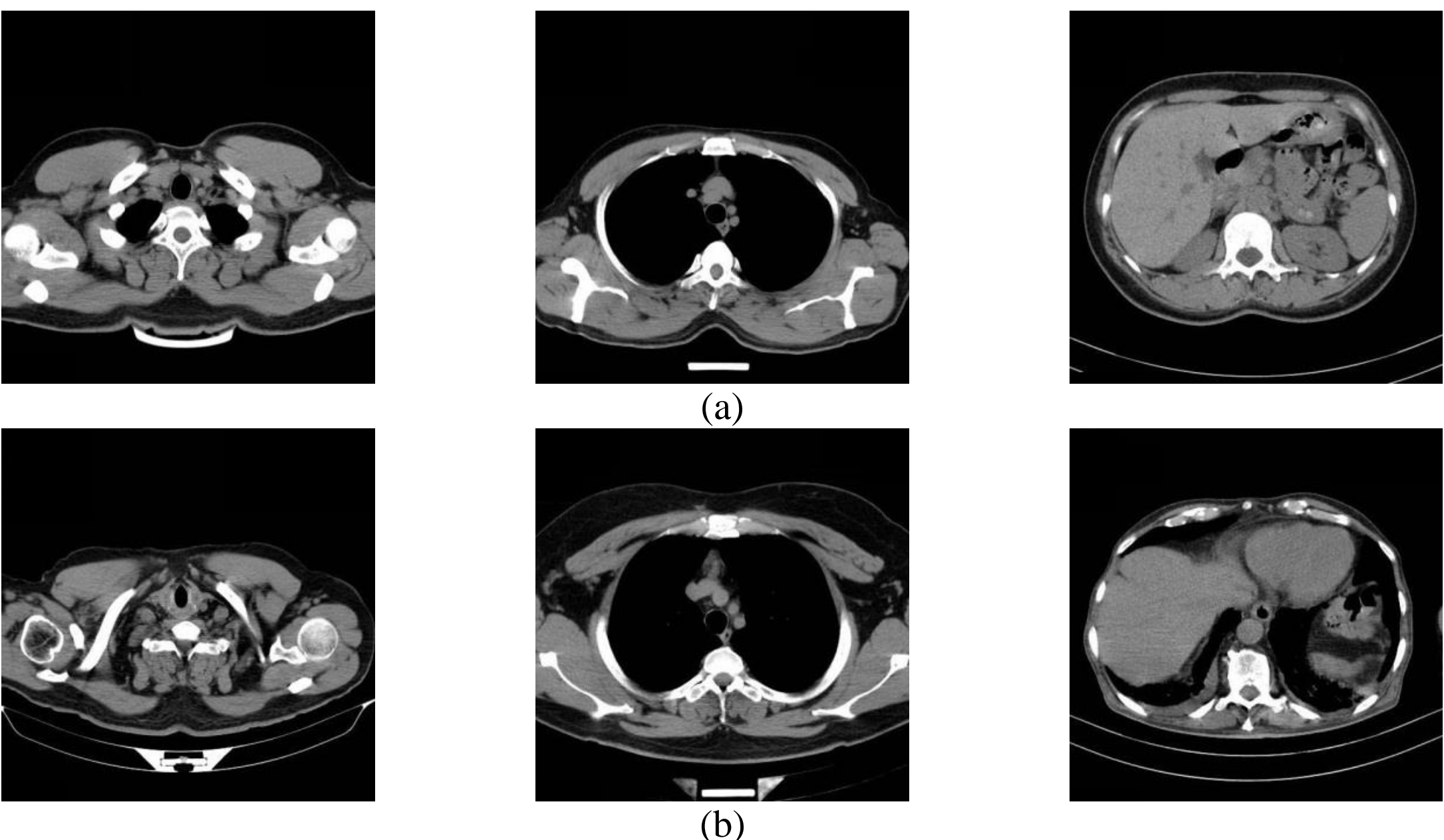

Figure 1. Typical CT scan images used in this work: (a) healthy subject, (b) COVID-19 infected patient.

# 4 Proposed method

Despite the fact that the new coronavirus is spreading quickly worldwide, the amount of available labelled data for diagnosing the disease is limited. Also, due to the rapid growth of this virus, the ability to label data quickly is desirable. Therefore, this study takes a semi-supervised approach to train the deep neural network for diagnosing coronavirus with high accuracy where only 10% of the training dataset is labelled. The proposed method has *two* phases. They are explained clearly in the following sections.

## 4.1 First phase: unsupervised training

The objective of the first phase is to learn the underlying distribution of dataset samples achieved using unsupervised training [43] of GAN model using unlabelled data. The reason for using GAN is its ability to learn the training data distribution. Upon completion of the first phase, the GAN networks parameters are trained such that the *generator* is capable of producing high quality fake images and the *discriminator* is capable of distinguishing fake image from the real ones. This means that the *discriminator* has learned to identify the valid CT scan image. The output of the first phase is the trained parameters of *discriminator* which are used as a starting point for the second phase.

## 4.2 Second phase: supervised training

The objective of the second phase is to fine tune the *discriminator* parameters such that it can classify CT scan images as COVID or healthy. To this end, the labelled data are used in supervised manner [44] to train the *discriminator*. It is worth noting that the *generator* is not needed in the second phase. The two phases of our method are depicted in Figure 2. As can be seen, the *discriminator* output in the first phase is the probability of being real given the input image while the output in the second phase is probability of being COVID infected. Each sample is classified as COVID if p(COVID) > 0.5, otherwise it is classified as Healthy. The advantage of probabilistic output is that it can also be used as an uncertainty measure. The more the *discriminator* output is close to 0.5, the higher the uncertainty about the sample class (Health/COVID) is. Hence, looking at the *discriminator* output, the human expert gets an insight on how much the classifier output can be trusted. In case the classification has low confidence (p(COVID)~0.5), the human expert can ask for another human/classifier to confirm the class of the test images.

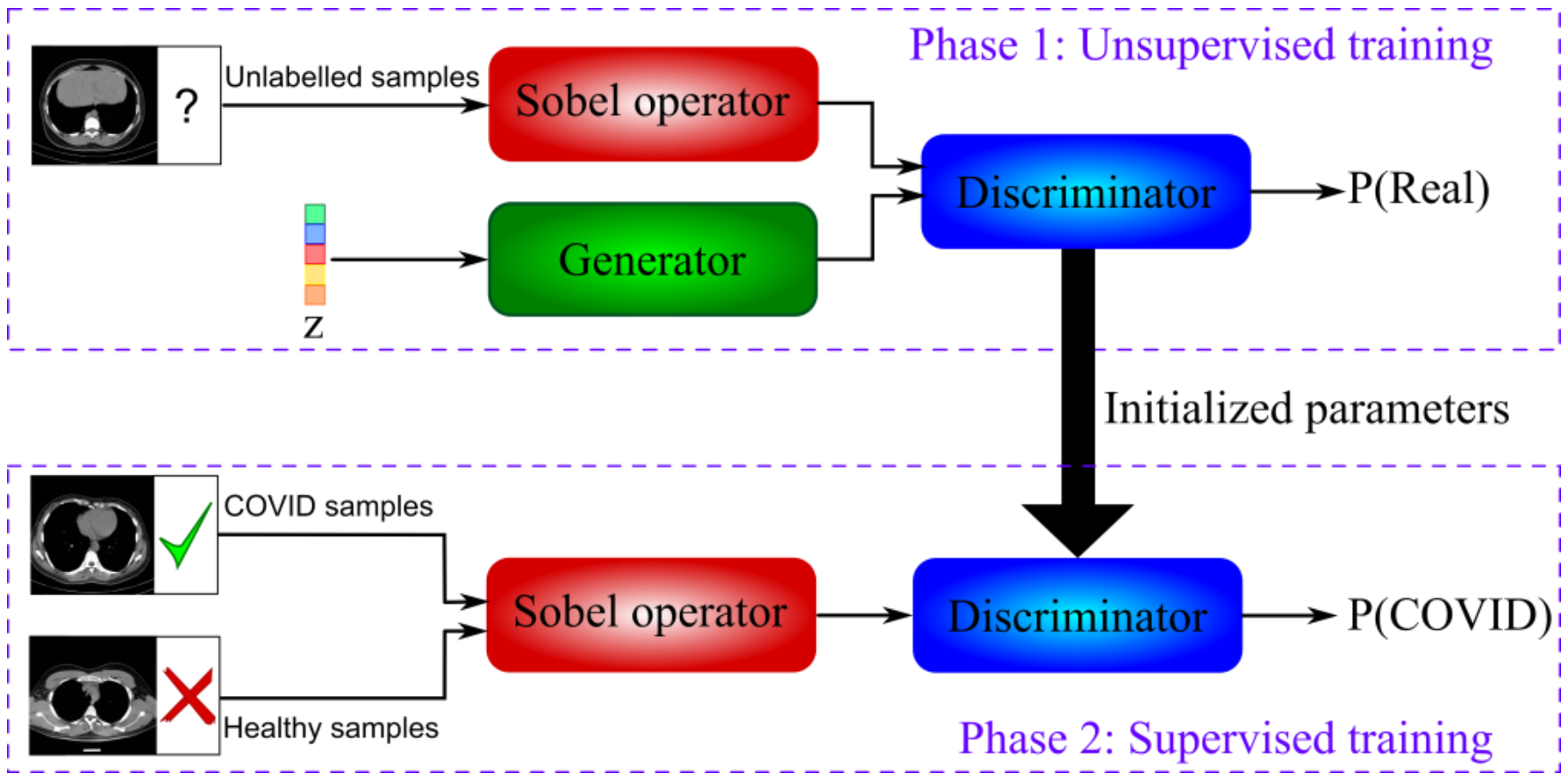

Figure 2. Schematic illustration of the proposed SCLLD model.

## 4.3 GAN architecture

The *generator* architecture [41] is shown in Figure 3. The input to the *generator* is a (Gaussian) noise vector which is fed to multiple convolutional layers. These layers convert the noise vector to synthetic CT scan images. The *discriminator* architecture is depicted in Figure 4. The input to *discriminator* is a two-dimensional image which could be a real sample or a fake one produced by the *generator.* During the forward pass of the *discriminator*, the input image is reduced to a scalar value which denotes the *discriminator* judgment about the received input sample. In the first phase, the ideal behaviour of the *discriminator* is to provide a higher probability of being real (p(real)) for real images and a lower probability for fake ones. Since the *discriminator* is competing against the *generator* (usually after the successful training), the probability provided by the *discriminator* is around 0.5 for both real and fake samples. This stems from the fact that the quality of the *generator's* fake samples is high enough to deceive the *discriminator*. The desired behaviour of the *discriminator* in the second phase is to distinguish between COVID and healthy images.

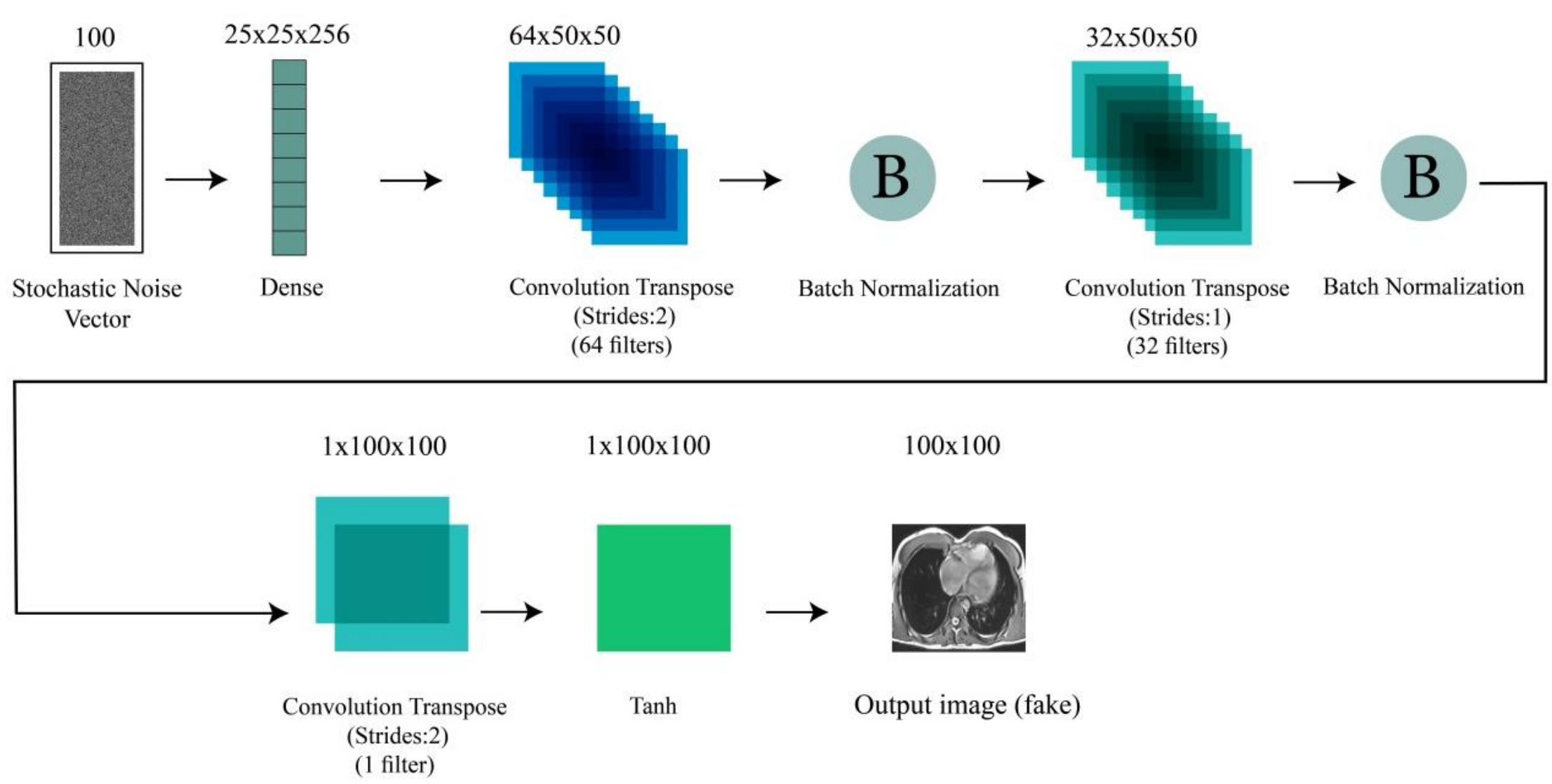

Figure 3. Network structure of *generator* model used in this work.

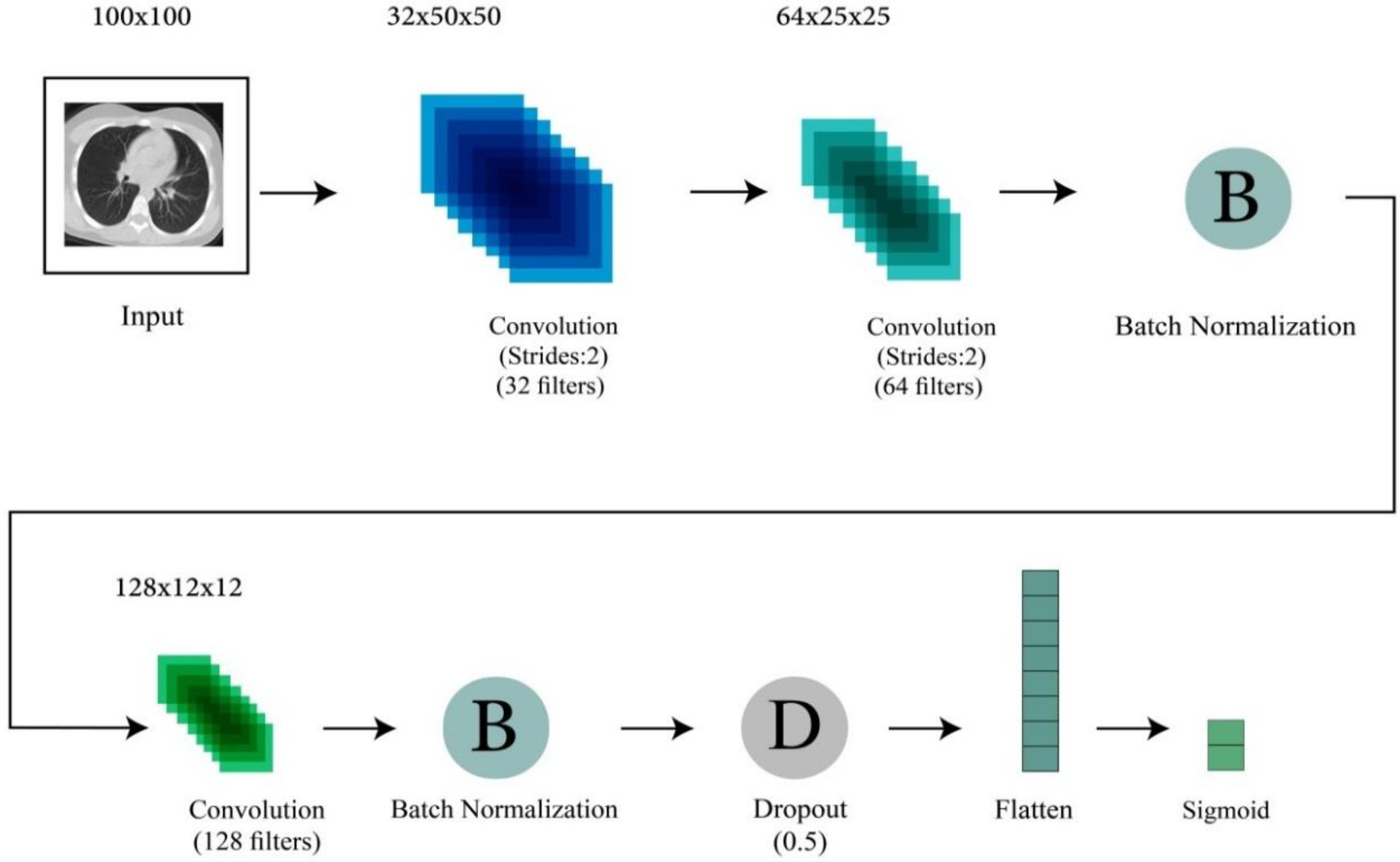

Figure 4. Network structure of *discriminator* model used in this work.

## 4.4 Dataset management

Our dataset contains 10000 CT scan images which are divided into 80% training (8000 images) and 20% testing (2000 images). The test data are all labelled. Moreover, 10% (800 images) of the training data is labelled. For validation, 20% (160 images) of the labelled training data is used.

To improve the performance, we pre-process the available data. The edge detection by Sobel operator is applied to the data to further improve the feature extraction in GAN.

## 4.5 Our method steps

The main steps of our method are illustrated in Figure 5. In the first step, the dataset is partitioned to training, testing, and validation as explained in section 4.4. In step two, 10% of the training samples and entire testing samples are labelled, manually. Before any pre-processing, in the third step, the Sobel technique is used to detect the edges of the images. Labelled and unlabelled data are pre-processed in step four. Initial pre-processing includes resizing of images to 100*100 and normalizing them. The *normalization* maps the intensity of all image pixels to interval [0, 1]. The fifth step involves the creation of GAN *generator* and *discriminator* networks' structures that are presented in Figure 3 and Figure 4, respectively. The unsupervised training of sixth step is exactly the same as standard training procedure of GAN. The *generator* is trained such that it deceives the *discriminator* with high-quality fake images. The *discriminator* is trained in an unsupervised way by using the unlabelled data to distinguish between real and fake images. In the seventh step, the *discriminator* (with parameters initialized in step 6) is trained in a supervised manner to gain the ability to classify input images as healthy or COVID patients. To this end, the *discriminator* last layer activation provides the probability of each image of being COVID. In this step, the labelled data comes into play. Although the number of labelled data is small, the *discriminator* is able to distinguish COVID and healthy samples since its parameters have already been initialized to sensible values in the previous step. Finally, the trained *discriminator* is used to classify the test data in step eight.

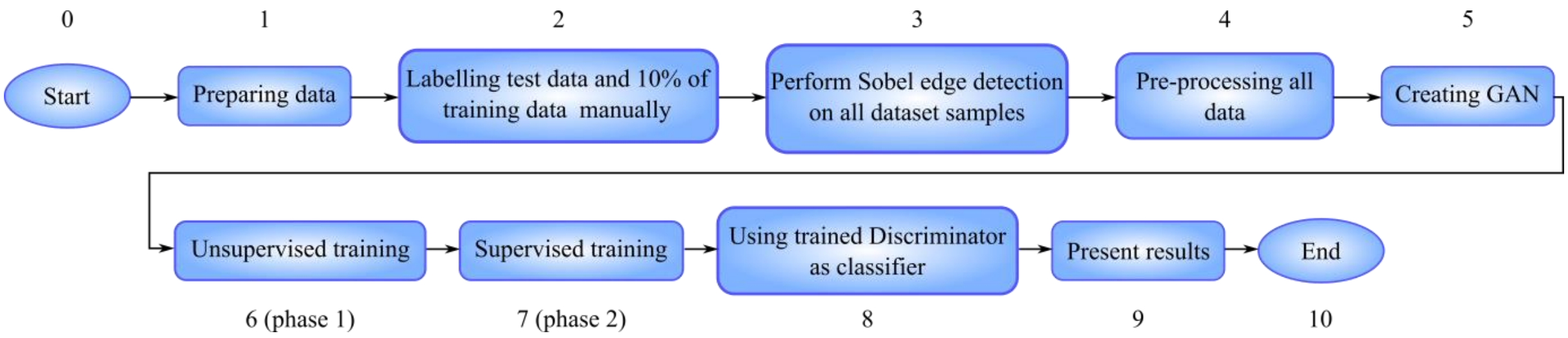

Figure 5. The proposed method steps

## 4.6 Implementation details

The proposed method has been implemented in Python using Keras library, which runs on top of TensorFlow. The experiments have been run on a PC with GFORCE GTX 950 GPU and 16GB of RAM. Hyperparameters of the method are explained in Table 2.

Table 2. Parameters of the proposed method

| | |
|---|---|
| Size of the convolution kernels | $3 \times 3$ |
| Generator input noise vector length | 100 |
| Batch size | 32 |
| Number of iterations | 3500, 4000, and 4500 |
| Optimizer method | Adam |
| Adam parameters | $\beta_1 = 0.9, \beta_2 = 0.999$ |
| Generator and discriminator loss | Binary Cross Entropy (BCE) |
| Generator learning rate | 1e-3 |
| Discriminator learning rate | 1e-3 |
| Convolutional layers activation function | Leaky ReLU ($\alpha = 0.01$) |
| Edge detection algorithm | Sobel with kernel size=3 |
| Dropout probability | 0.5 |

# 5 Results

In this section, the experimental results of GAN and SCLLD methods for detection of COVID-19 patients based on CT scan data are presented. For SCLLD method, we provided three loss plots at iterations 3500, 4000, and 4500 averaged on five number of runs with 95% confidence intervals. These plots are shown in parts a-c of Figures 6-8. Part d of these figures show ROC plots for SCLLD method.

Moreover, we have compared the methods based on criteria such as accuracy, precision, sensitivity, recall, specificity, F1-score, and AUC. These criteria are computed according to the following equations [13]:

$$\text{Accuracy} = \frac{\text{TP} + \text{TN}}{\text{FP} + \text{FN} + \text{TP} + \text{TN}}, \tag{8}$$

$$\text{Sensitivity or Recall} = \frac{\text{TP}}{\text{TP} + \text{FN}}, \tag{9}$$

$$\text{Precision} = \frac{\text{TP}}{\text{TP} + \text{FP}}, \tag{10}$$

$$\text{F1} - \text{Score} = \frac{2\text{TP}}{2\text{TP} + \text{FP} + \text{FN}}, \tag{11}$$

$$\text{Specificity} = \frac{\text{TN}}{\text{TN} + \text{FP}}. \quad (12)$$

The experimental results for SCLLD method are presented in section 5.1. We have also summarised the results in Table 3.

## 5.1 SCLLD results on our dataset

The results of experiments at iterations 3500, 4000, and 4500 of SCLLD method are presented in Figures 6-8. The proposed method has *two* training phases: unsupervised and supervised. The loss plots in parts a and b of Figure 6 belong to the unsupervised training phase while the loss plot in part c is related to supervised training. Part d of the figure illustrates the ROC plot of trained SCLLD. The structure of plots in figures 7 and 8 are similar to the one explained in Figure 6. The only difference is the iterations (4000 and 4500) at which the plots are presented. The motivation behind plotting the results at multiple points during training is the investigation to the increasing number of iterations. This investigation is important since at some point during the training, the performance of SCLLD may degrade. Therefore, it is common practice to diagnose the SCLLD performance and stop the training when the performance of the model does not improve any more.

It can be noted that, parts a and b of Figures 6-8 reveal that using Sobel operator has accelerated the training pace. Therefore, the best performance is achieved at iteration 3500. Beyond that iteration, the model has started to degrade quickly, leading to considerable loss value at iteration 4500 (part b of figure 8).

The ROC plots at different iterations of SCLLD method match the accuracy results reported in Table 3. Considering that the best performance is achieved at iteration 3500, it makes sense that ROC plots at this iteration reach value of 1.0 faster compared to their counterparts at iterations 4000 and 4500. For SCLLD, the accuracy decreases at iteration 4000 but increases slightly at iteration 4500. That is why ROC at iteration 4500 increases faster as compared to ROC at iteration 4000. An example of the CT images, result of applying Soble filter on them and the final results of GRAD-CAM are shown in Figure 9. Grad-CAM is a generalization of the Class Activation Mapping. It does not needs re-training and can be applied broadly to any CNN-based architectures [45].

### 5.1.1 Sensitivity analysis regarding labelled training data size

Based on the existing literature, GAN is good at capturing the underlying distribution of the training dataset [42]. However, the limited number of available labelled data might hurt the classification performance. This is the motivation for our semi-supervised approach SCLLD. In this section, we evaluated the performance of our method for different number of labelled data. This experiment reveals how much our semi-supervised approach can tolerate limited labelled data. The results for increasing sample size of training labelled data (1% to 10%) are presented in Table 4. It is clear that even in a semi-supervised setting, the amount of labelled data cannot be less than a certain threshold, otherwise the performance drops dramatically. This scenario is observed if the labelled data falls less than 3%. For labelled data size above 6%, the performance metrics are similar and the best results are achieved when 8% of the data are labelled. Theoretically, increasing the training data should lead to better performance. However, it is customary to train deep neural networks using mini-batch sizes smaller than the whole dataset. The benefit of mini-batch training is twofold [46]. First, smaller batch size reduces the memory demand for one step of training. Second, smaller batch size results in noisy gradients which has regularization effect preventing over-fitting. The noisy gradients lead to stochasticity during the learning process which is why the accuracy values in Table 4 do not necessarily increase with larger labelled data size.

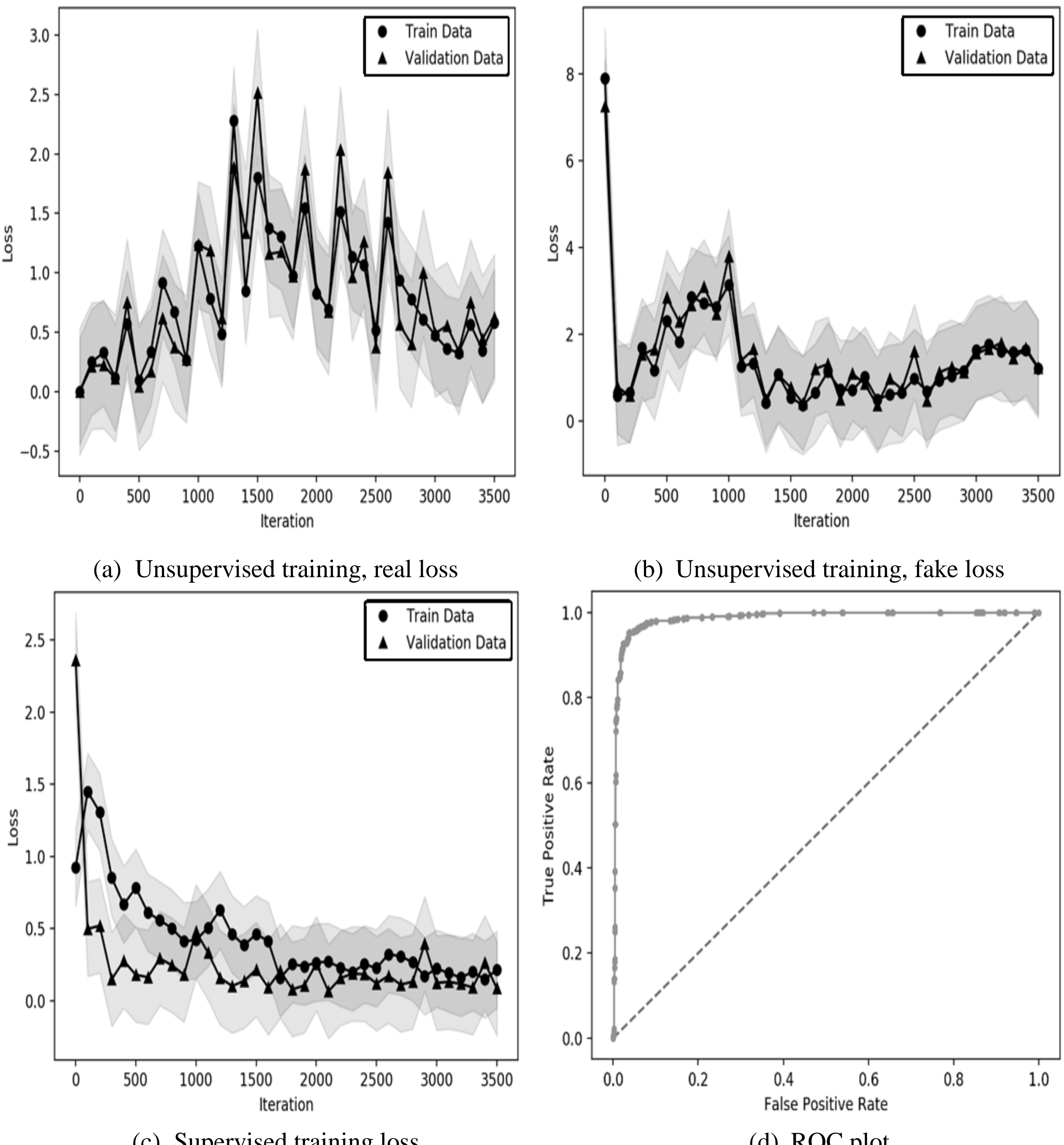

(a) Unsupervised training, real loss

(b) Unsupervised training, fake loss

(c) Supervised training loss

(d) ROC plot

Figure 6. SCLLD method results at iteration 3500: (a) loss plot on real data during unsupervised training, (b) loss plot on fake data (Generator output) during unsupervised training, (c) Loss plot for labelled data during supervised training and (d) ROC plot for classification of labelled data.

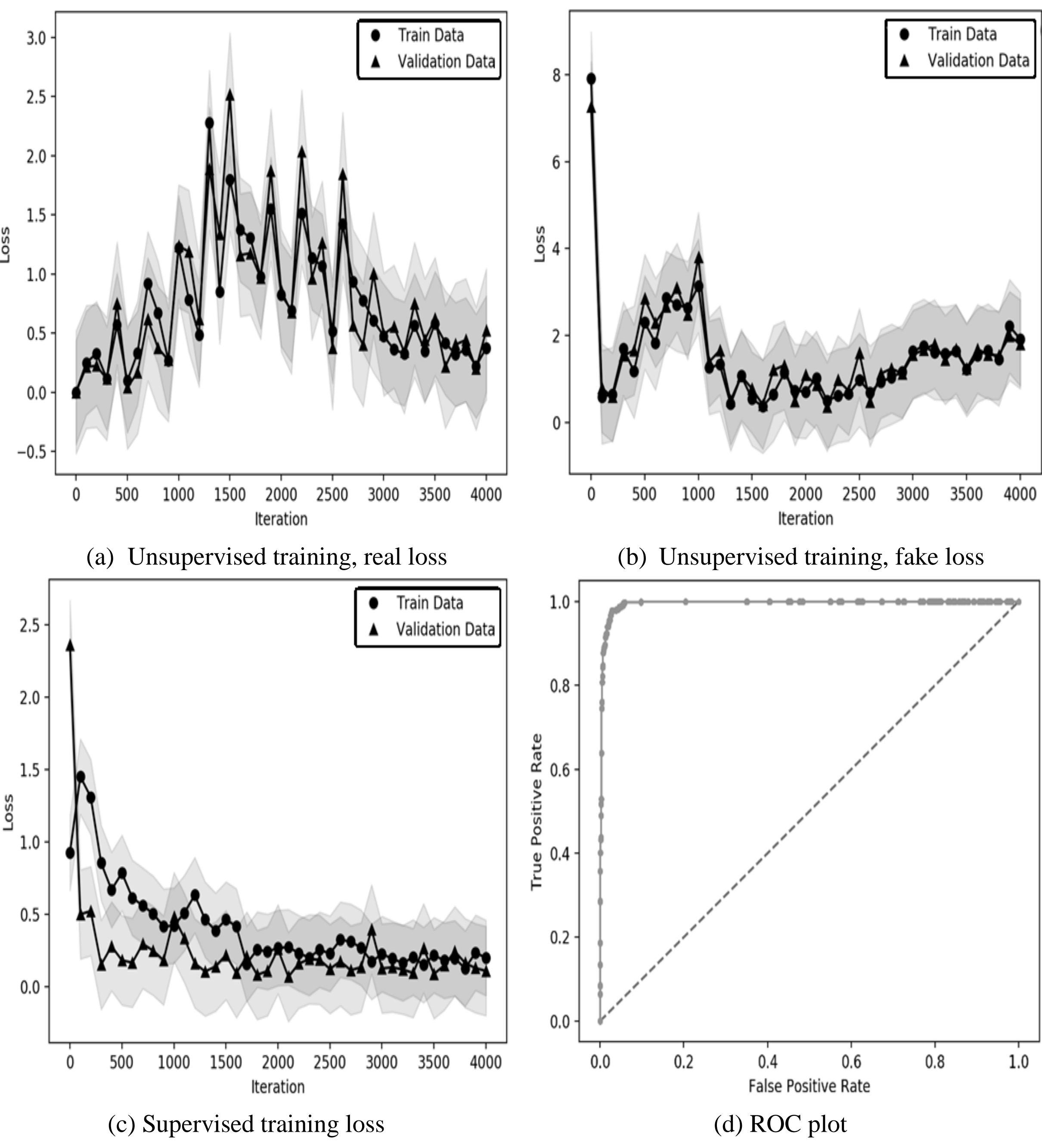

(a) Unsupervised training, real loss

(b) Unsupervised training, fake loss

(c) Supervised training loss

(d) ROC plot

Figure 7. SCLLD method results at iteration 4000: (a) loss plot on real data during unsupervised training, (b) loss plot on fake data (Generator output) during unsupervised training, (c) Loss plot for labelled data during supervised training and (d) ROC plot for classification of labelled data.

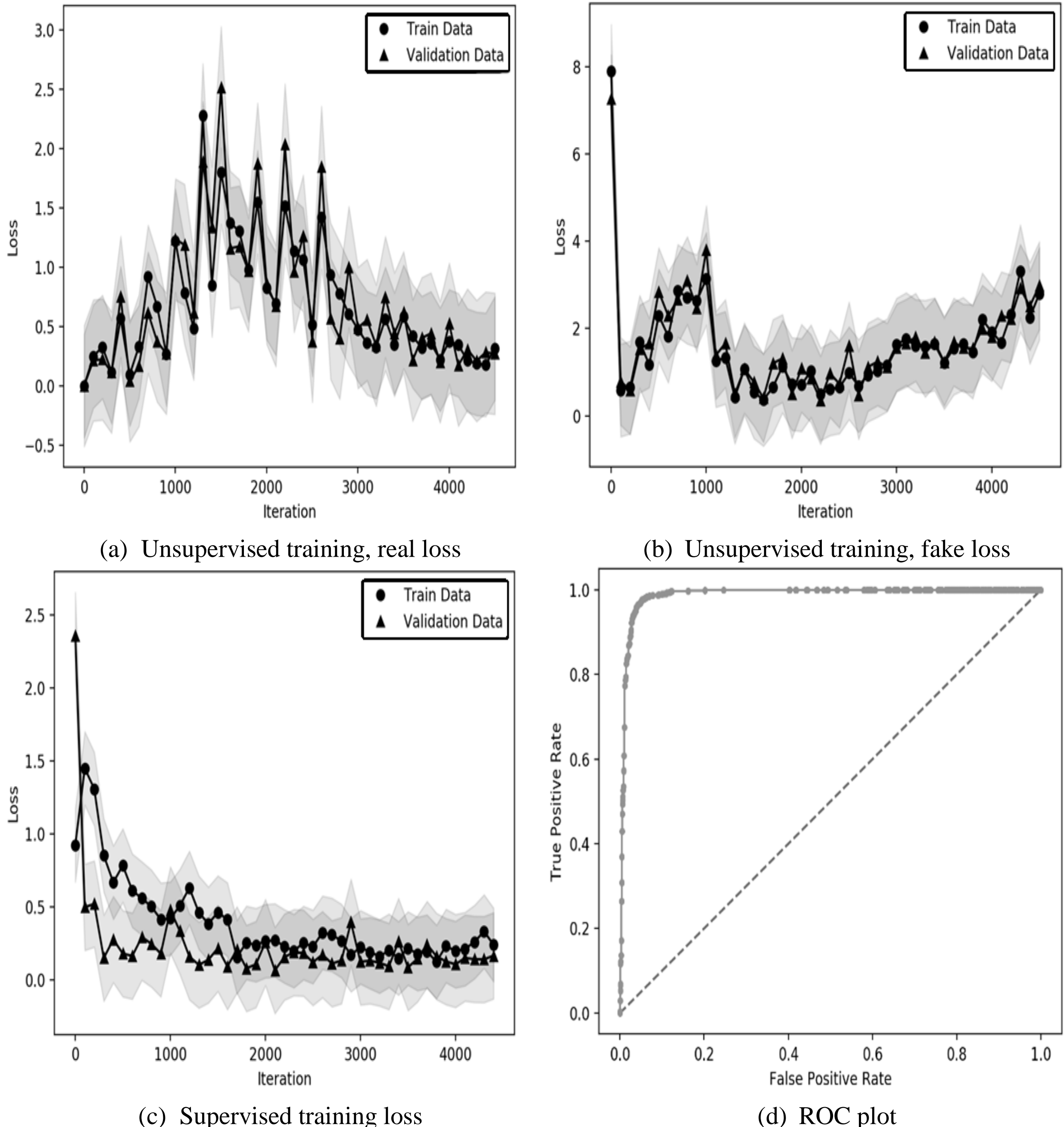

(a) Unsupervised training, real loss

(b) Unsupervised training, fake loss

(c) Supervised training loss

(d) ROC plot

Figure 8. SCLLD method results at iteration 4500: (a) loss plot on real data during unsupervised training, (b) loss plot on fake data (Generator output) during unsupervised training, (c) Loss plot for labelled data during supervised training and (d) ROC plot for classification of labelled data.

Table 3. Comparison results between methods of GAN and SCLLD through different evaluation metrics.

| Methods | Number of Iterations | Accuracy (%) | Precision (%) | Recall (%) | Specificity (%) | F1-score (%) | Loss | AUC (%) |
|---|---|---|---|---|---|---|---|---|
| GAN | 3500 | 98.42 | 99.39 | 97.39 | 99.42 | 98.38 | 0.2567 | 98.4061 |
| | 4000 | 98.37 | 100 | 96.69 | 100 | 98.32 | 0.2856 | 98.3466 |
| | 4500 | 97.09 | 100 | 94.09 | 100 | 96.95 | 0.2205 | 97.0440 |
| SCLLD | 3500 | 99.61 | 99.8 | 99.4 | 99.81 | 99.6 | 0.2548 | 99.6023 |
| | 4000 | 98.77 | 100 | 97.49 | 100 | 98.73 | 0.1570 | 98.7474 |
| | 4500 | 98.96 | 97.94 | 100 | 97.96 | 98.96 | 0.3853 | 98.9805 |

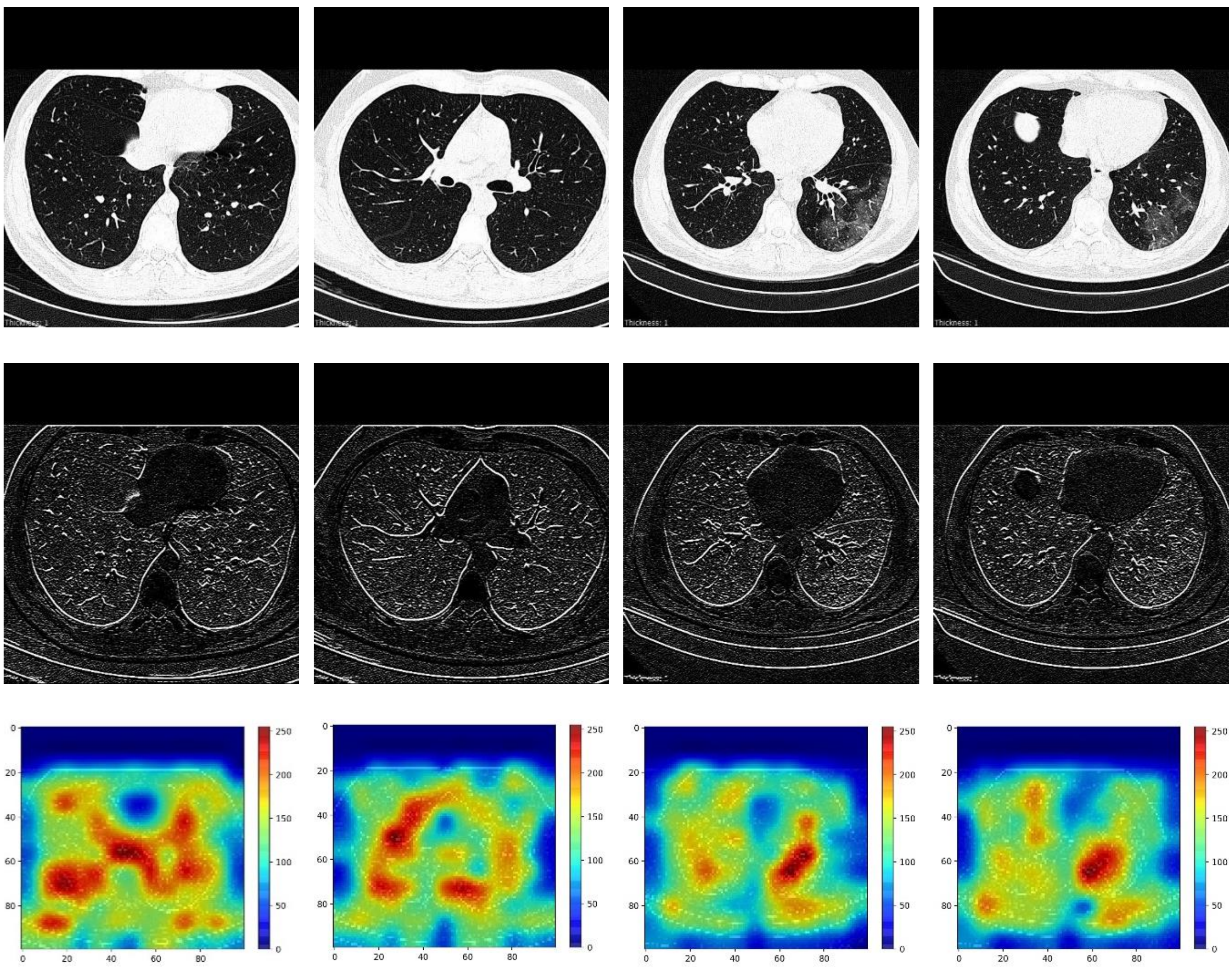

Figure 9. Examples of the CT images, result of applying Soble filter on them and the final results of GRAD-CAM are shown in the first, second and third rows, respectively. The first and second columns are the images of normal cases while the third and fourth columns are the images of sick cases.

As can be seen in Table 4, the total training time is not a function of labelled data size. This is due to the fact that regardless of being labelled or not, all samples are eventually used during the supervised or semi-supervised training. Hence, the total cost of training is determined by the total number of labelled/unlabelled samples.

Table 4. Effect of labelled training data size on SCLLD method performance within 4000 iterations.

| **Amount of labelled training data (%)** | **Accuracy (%)** | **Precision (%)** | **Recall (%)** | **Specificity (%)** | **F1-score (%)** | **Loss** | **AUC (%)** | **Total training time (minute)** |
|---|---|---|---|---|---|---|---|---|
| 1 | 52.37 | 100 | 3.21 | 100 | 6.22 | 2.0755 | 51.6 | 0:26:52.736555 |
| 2 | 64.6 | 83.18 | 35.17 | 93.11 | 49.44 | 1.0012 | 64.14 | 0:26:52.654560 |
| 3 | 93.54 | 89.52 | 98.4 | 88.83 | 93.75 | 1.2660 | 93.62 | 0:26:57.976252 |
| 4 | 96.15 | 98.52 | 93.59 | 98.64 | 95.99 | 0.2459 | 96.11 | 0:27:06.940879 |
| 5 | 95.36 | 91.7 | 99.6 | 91.26 | 95.49 | 1.0138 | 95.43 | 0:26:55.096258 |
| 6 | 99.26 | 98.91 | 99.6 | 98.93 | 99.25 | 0.2071 | 99.27 | 0:26:56.961483 |
| 7 | 99.56 | 99.8 | 99.3 | 99.81 | 99.55 | 0.0963 | 99.55 | 0:26:47.955172 |
| 8 | **99.95** | 100 | 99.9 | 100 | 99.95 | 0.1823 | 99.95 | 0:26:57.434292 |
| 9 | 99.41 | 99.7 | 99.1 | 99.71 | 99.4 | 0.1091 | 99.4 | 0:26:56.027641 |
| 10 | 99.36 | 98.81 | 99.9 | 98.83 | 99.35 | 0.2391 | 99.37 | 0:26:54.960978 |

## 5.2 Evaluation of Convolution Neural Network on our dataset

To highlight the role of using unlabelled data in our method, we have also trained a Convolutional Neural Network (CNN) on labelled data in a purely supervised manner. The structure of CNN is the same as our GAN *discriminator*. According to Figure 10, it is possible to compare the loss of CNN and our proposed method during the train and validation phases. According to this figure, it is clear that although the loss function of our proposed method is worse than CNN during the training phase, it shows better performance during the validation phase.

The accuracy on training data and validation data during the training of CNN is shown in Figure 11. Looking at the accuracy plot, it is evident that the training has been stopped early by Keras library. The reason for early stopping is the fact that the CNN performance has started to degrade after iteration 11.

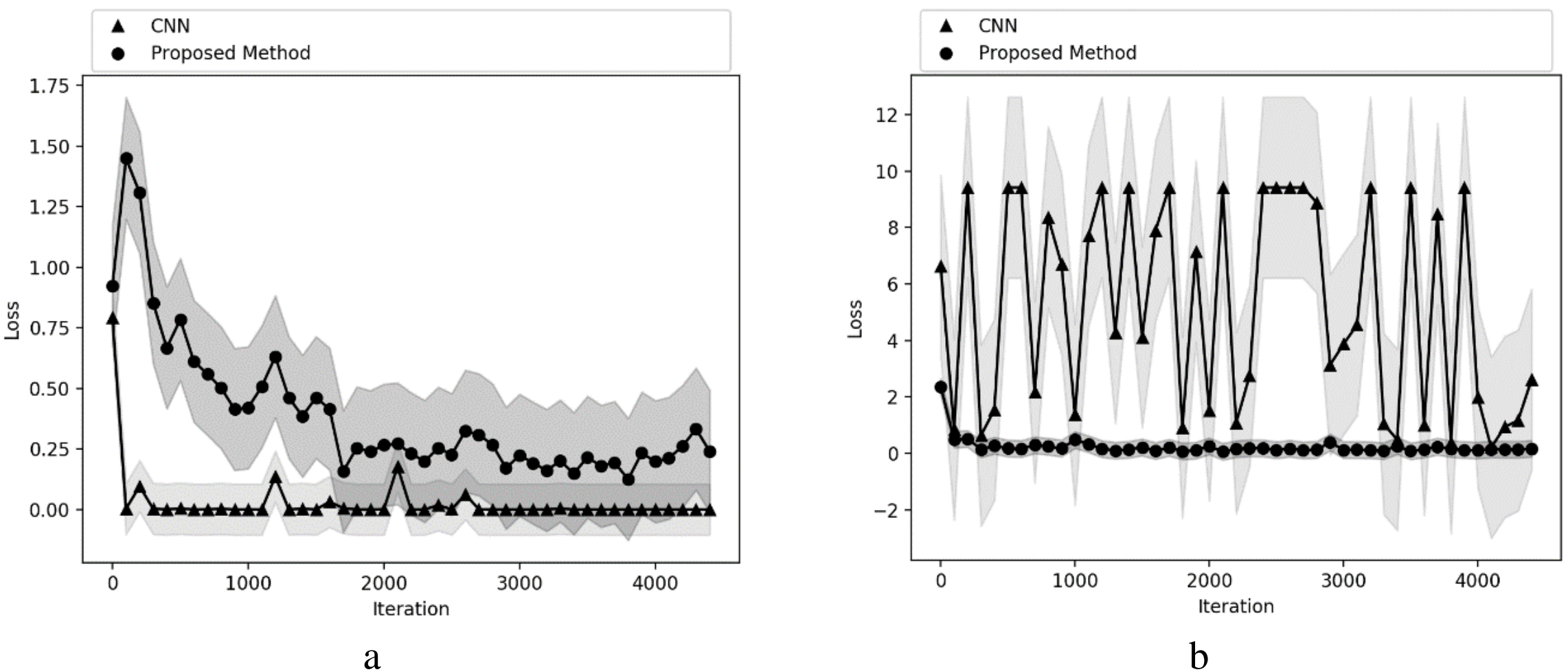

Figure 10. Loss plot of CNN and the Proposed Method during a) training phase, b) validation phase

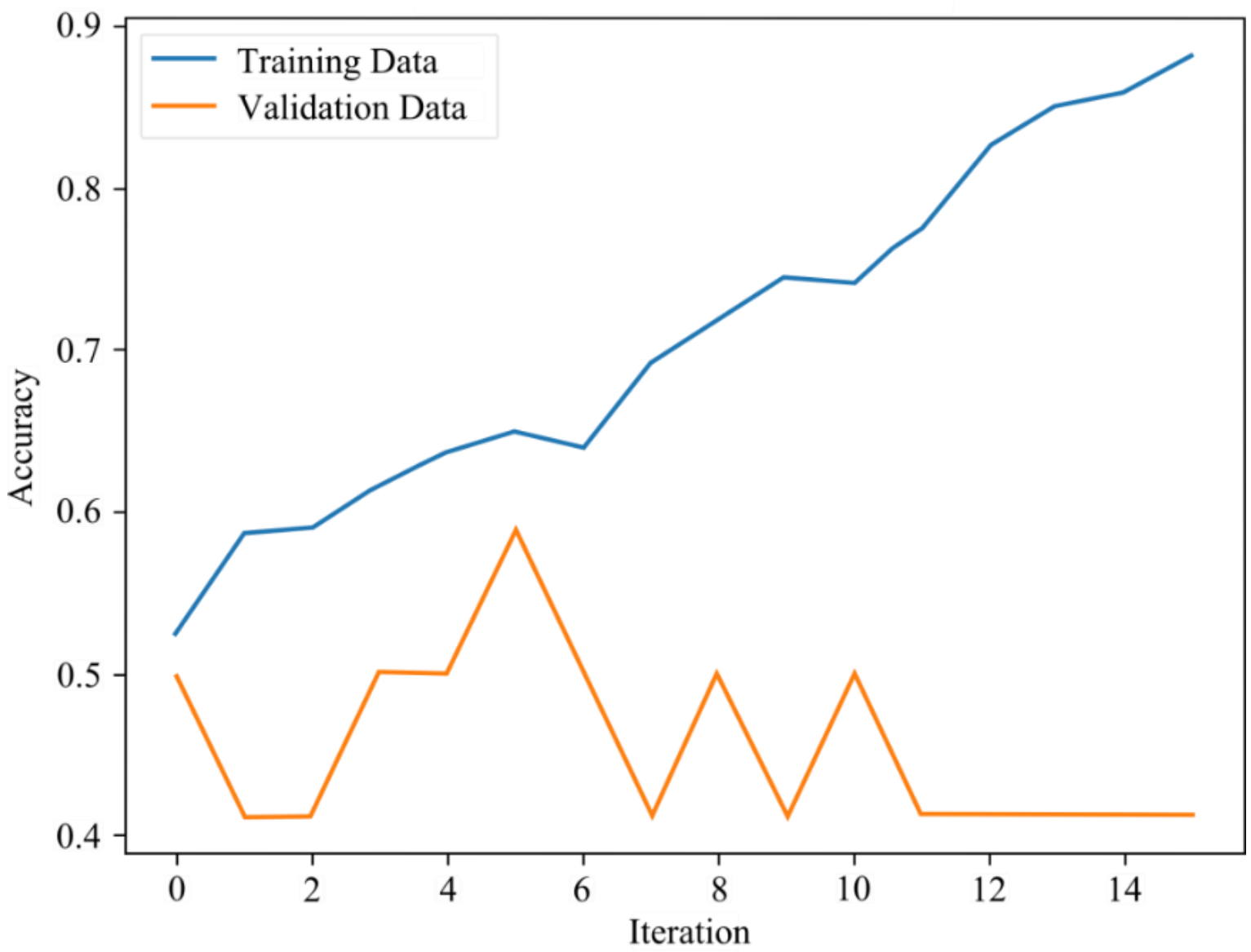

Figure 11. The accuracy plot during the training of CNN on labelled data

The performance of the trained CNN on test data is presented in Table 5. It is clear that CNN training has suffered from the lack of sufficient amount of labelled data. While CNN requires a huge amount of labelled training data in order to unleash its representation power; our method can enjoy unlabelled data to reach significantly better performance even if labelled data are limited.

Table 5. Performance metrics of trained CNN on test data

| Accuracy (%) | Precision (%) | Recall (%) | Specificity (%) | F1-score (%) | AUC (%) | Total training time (minute) |
|---|---|---|---|---|---|---|
| 69.87 | 63.00 | 94.00 | 46.40 | 75.00 | 70.24 | 0:00:09.940334 |

## 5.3 Evaluation of Gaussian Process on our dataset

To gain a better understanding of our results, we have also compared our proposed method with a Gaussian process (GP) approach which is trained in supervised manner with our dataset. The advantage of GP is that it can represent any dataset. The drawback is its high computational complexity, which is of the order $O(N^3)$ [47] for N training samples. Considering that the amount of labelled data in our dataset is small, GP is an ideal choice since (i) the computation overhead is limited as the amount of available labelled data is limited. (ii) GP provides an estimate of the uncertainty in portions of the sample space where it has not seen enough data so the user knows where the GP output can be trusted.

As can be seen in Table 6, despite its power, GP exhibited poor performance due to lack of sufficient labelled data. We can observe that even robust supervised methods cannot deal with limited labelled data. This clearly shows the importance of semi-supervised learning methods. The ROC diagram of GP experiment on our dataset is also presented in Figure 12.

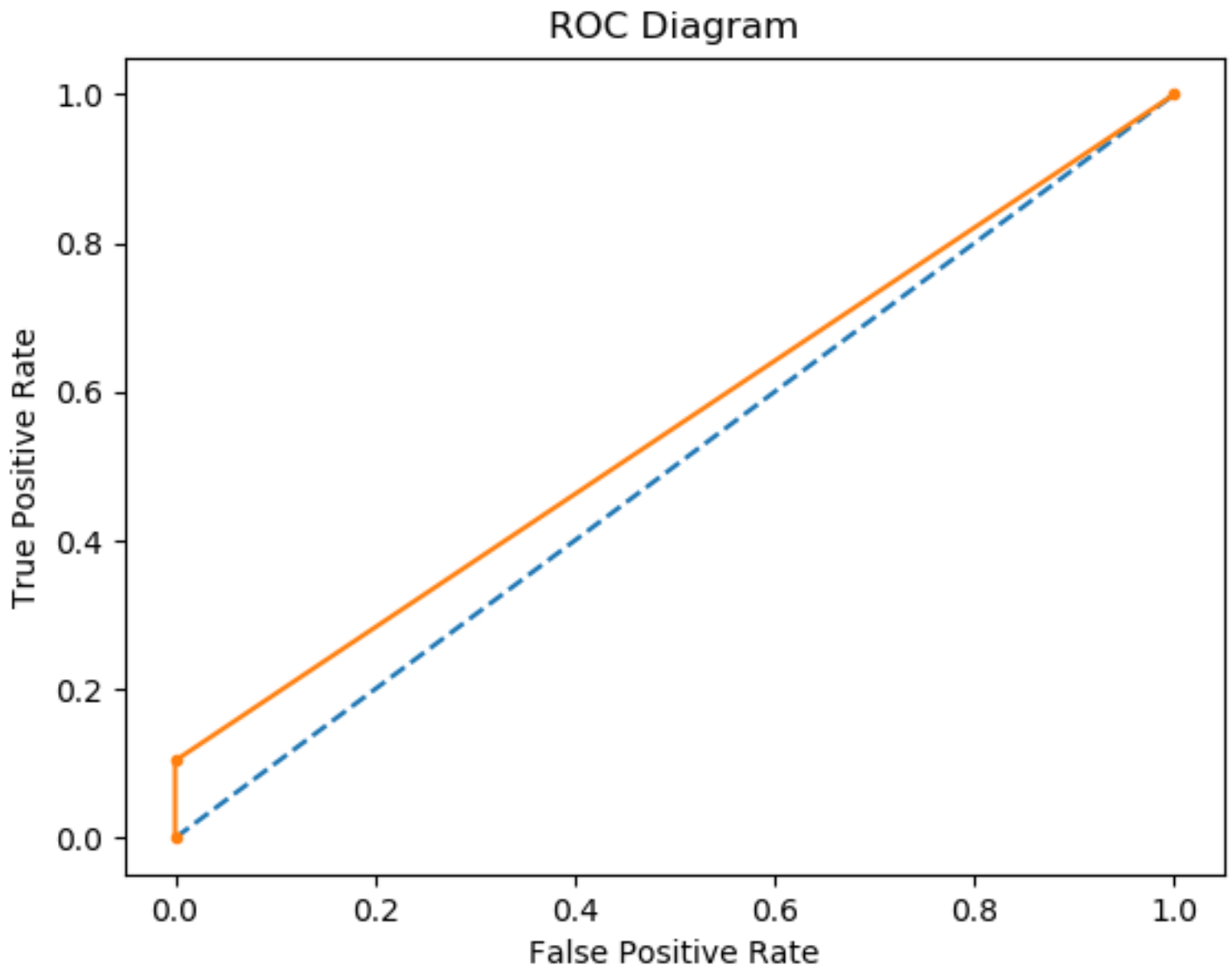

Figure 12. ROC diagram obtained using our method with our dataset and GP.

Table 6. Performance of GP with our dataset.

| Accuracy (%) | Precision (%) | Recall (%) | Specificity (%) | F1-score (%) | AUC (%) | Total training time (minute) |
|---|---|---|---|---|---|---|
| 54.48 | 52 | 100 | 10.38 | 68 | 55.19 | 0:16:52.914259 |

## 5.4 Evaluation of our method using a public dataset

To evaluate the generalization of our method, we have evaluated the performance of our model using the publicly available Kaggle dataset [13]. The dataset belongs to China National Centre for Bio-information with 8535 positive and 9430 negative CT scan samples. It can be noted from the presented results in Table 7 that our method has shown reasonable performance. Hence, our method is not limited to a specific dataset.

Table 7. Results obtained using GAN and SCLLD methods with different evaluation metrics after 4000 training iterations with 80% of Kaggle dataset [13] for training and 20% for testing.

| Methods | Accuracy (%) | Precision (%) | Recall (%) | Specificity (%) | F1-score (%) | Loss | AUC (%) | Total training time (minute) |
|---|---|---|---|---|---|---|---|---|
| GAN | 89.98 | 98.77 | 79.91 | 99.1 | 88.34 | 0.3669 | 89.50 | 0:26:56.824186 |
| SCLLD | 96.74 | 95.17 | 98.13 | 95.49 | 96.63 | 0.1977 | 96.81 | 0:26:59.223519 |

# 6 Discussion and Conclusion

COVID-19 pandemic has led to more than 1 million deaths and is spreading faster than expected. Currently, early detection of virus is of the highest importance which is a challenging task. Moreover, the virus detection test is expensive and not robust. Hence, chest x-ray and CT scans may be the useful tool to detect COVID-19 at an early stage. In addition to these detection techniques, researchers are utilizing AI methods to accelerate the detection and treatment of COVID-19 infected patients. This is the motivation behind our proposed method based on DL using a semi-supervised method.

Our objective is to classify healthy and COVID-19 infected patients accurately using chest CT scan images. We have used the GAN method to deal with limited labelled data. To this end, GAN *discriminator* network parameters are initialized using unlabelled data. Next, the network parameters are fine-tuned with small number of labelled data. The proposed approach is further improved by pre-processing the input images using Sobel edge detection. After the training, the GAN *discriminator* is used to classify the test data. Using the probabilistic *discriminator* output, our system provides the human experts with an uncertainty measure about its decision. The experimental results reveal the superiority of SCLLD method as compared to the GAN method alone (Table 3).

To evaluate our SCLLD against other supervised methods, we have compared SCLLD performance with our dataset and other methods using their datasets (Table 8). Testing on different datasets is necessary since our dataset has limited labelled data which can hurt the performance of supervised methods. This comparison reveals that while supervised learners are crippled in the absence of enough labelled data, our method is capable of dealing with limited labelled data and learning from unlabelled ones reaching performance which is on par with supervised methods. This is clearly the advantage of our method, which is due to its semi-supervised nature.

The drawback of our method is its computation overhead. In this work, we have used two training phases. In the first phase, standard GAN training is executed, which involves training of *generator* and *discriminator* together. The first source of overhead is *generator* training since when training is done, it is no longer needed. In the second phase of training, *discriminator* is trained using labelled data. Hence, the second source of overhead is training of *discriminator* twice (unsupervised and supervised). We may note that the final output of our method is the trained *discriminator*.

In future work, we investigate the effect of training the generator using feature matching [48]. Moreover, it is interesting to explore the effect of using soft labels [48] during the unsupervised training phase of GAN. To this end, hard labels (zero and one) corresponding to fake and real samples are replaced with a random number between [0, 0.3] and [0.7, 1], respectively. Soft labels usually lead to stable training.

Table 8. Summary of comparison with other automated DL methods developed to detect COVID-19 with CT images.

| Authors | DL method | Training approach | Number of images | Performance |
| --- | --- | --- | --- | --- |
| Ardakani et al. [32] | ResNet-101 | Supervised | 1020 CT | AUC: 0.994<br>Sensitivity: 100%<br>Specificity: 99.02%<br>Accuracy: 99.51% |
| Li et al. [18] | Resnet50 | Supervised | 4356 CT | Specificity: 92%<br>Sensitivity: 87%<br>AUC: 0.95 |
| Wang et al. [19] | CNN | Supervised | 453 CT | Sensitivity: 74%<br>Specificity: 67%<br>Accuracy: 73.1% |
| Gozes et al. [20] | Resnet-50 | Supervised | 6150 CT | Specificity: 92.2%<br>Sensitivity: 98.2%<br>AUC: 0.996 |
| Hemdan et al. [21] | Google MobileNet and modified VGG19 | Supervised | 50 X-ray | F1 score: 91% |
| Apostolopoulos et al. [23] | CNN with Transfer Learning | Supervised | 1427 X-ray | Specificity: 96.46%<br>Sensitivity: 98.66%<br>Accuracy: 96.78% |
| Butt et al. [24] | CNN | Supervised | 618 CT | Specificity: 92.2%,<br>Sensitivity: 98.2%<br>AUC: 0.996 |
| Song et al. [25] | Resnet50 | Supervised | 1485 CT | Sensitivity: 0.93<br>AUC: 0.99 |
| Sethy et al. [26] | SVM plus Resnet50 | Supervised | 50 X-ray | Kappa: 90.76%<br>F1 score: 91.41%<br>FPR : 95.52%<br>Accuracy: 95.38% |
| Narin et al. [27] | Inception-ResNetV2, InceptionV3 and ResNet50 | Supervised | 3141 X-ray | Accuracy: 98%. |
| Shi et al. [28] | VB-Net | Supervised | 196 CT | AUC: 0.89 |
| Proposed method | SCLLD | Semi-supervised | 10000 CT | Accuracy: 99.60%<br>Specificity: 99.80%<br>Sensitivity: 99.39%<br>AUC: 99.60% |

## Acknowledgement

This work was partly supported by the MINECO/ FEDER under the RTI2018-098913-B100, CV20-45250 and A-TIC-080-UGR18 projects.